\documentclass[a4paper, 12pt, english]{iopart} 

\usepackage[T1]{fontenc}
\usepackage{graphicx}
\usepackage[usenames,dvipsnames]{xcolor}
\usepackage[colorlinks=true, allcolors=blue]{hyperref}
\usepackage{siunitx}

\expandafter\let\csname equation*\endcsname\relax
\expandafter\let\csname endequation*\endcsname\relax
\usepackage{amsmath}
\usepackage{amsfonts}
\usepackage{amssymb}
\usepackage[inline]{enumitem}

\usepackage[usenames,dvipsnames]{xcolor}
\usepackage{framed}

\definecolor{cream}{rgb}{1.0, 0.99, 0.82}
\definecolor{celadon}{rgb}{0.67, 0.88, 0.69}
\definecolor{beaublue}{rgb}{0.74, 0.83, 0.9}
\definecolor{lightorange}{rgb}{1, 0.83, 0.56}
\definecolor{lightblue}{rgb}{0.7, 0.6, 1}

\definecolor{shadecolor}{rgb}{1.0, 0.99, 0.82}

\usepackage{soul}

\newcommand{\poisson}[2]{\mathcal{P}(#2|#1)}

\newcommand{\pon}{\alpha_{1}} 
\newcommand{\poff}{\beta_{1}} 
\newcommand{\roff}{r_{\beta}} 
\newcommand{\ron}{r_{\alpha}} 
\newcommand{\dpoff}{\beta_{d}} 
\newcommand{\dpon}{\alpha_{d}} 
\newcommand{\leg}{{l}}
\newcommand{\sd}{{\zeta}}  

\begin{document}

\title{Bayesian estimation of switching rates for blinking emitters}

\date{\today}

\author{Jemy~Geordy$^{1,2}$,
		Lachlan~J~Rogers$^{1,2}$,
		Cameron~M~Rogers$^3$,
		Thomas~Volz$^{1,2}$,
		Alexei~Gilchrist$^{1,2}$
}

\address{$^1$Department of Physics and Astronomy, Macquarie University, Sydney, NSW 2122, Australia.}
\address{$^2$ARC Centre of Excellence for Engineered Quantum Systems.}
\address{$^3$Launceston Church Grammar School, Mowbray Heights, Tasmania, 7248, Australia.}

\ead{lachlan.j.rogers@quantum.diamonds}

\begin{abstract}
Single quantum light-emitters are valuable resources for engineered quantum systems. 
They can function as robust single-photon generators, allow optical control of single spins, provide readout capabilities for atomic-scale sensors, and provide interfaces between stationary and flying qubits. 
Environmental factors can lead to single emitters exhibiting ``blinking'', whereby the fluorescence level switches between \textit{on} and \textit{off} states. 
Detailed characterisation of this blinking behaviour including determining the switching rates is often a powerful way to gain understanding about the underlying physical mechanisms.
While simple thresholds can be used to identify the \textit{on} and \textit{off} intervals and thus extract the rates from the time-series of counts for bright emitters with low background noise, such approaches become difficult for emitters fluorescing at low levels, high noise, or switching at fast rates. 
We develop a Bayesian approach capable of inferring switching rates directly from the time-series. 
This is able to deal with high levels of noise and fast switching in fluorescence traces. 
Moreover, the Bayesian inference also yields a robust picture of the parameter uncertainties, providing a benefit also for bright emitters in low-noise settings.
The technique can be adapted to identify the underlying states as well as extracting the rates of switching.
Finally, our method is applicable to a broad range of systems that show behaviour analogous to a single blinking emitter.
\end{abstract}

\maketitle

\section{Introduction}

A wide variety of natural and artificial systems exhibit an unwanted stochastic switching between states, sometimes called telegraph noise \cite{efros1997random}.
To remove this unwanted behaviour it is necessary to understand its causes, and to gain an understanding of the causes it is necessary to be able to measure the characteristics of the switching.
A prominent example of a system exhibiting this random switching of state is provided by quantum emitters where the behaviour is called \emph{blinking} due to large intermittent changes in the observed fluorescence.
Such emitters are valuable resources for engineered quantum systems, and the presence of blinking limits their applicability. 
If fact, most quantum light emitters exhibit intermittency in their fluorescence under continuous excitation, including 
diamond colour centres \cite{jantzen2016nanodiamonds, Neu_silicon, Carlo, Stefania, Wang}, 
quantum dots \cite{Nirmal_qdot, Pavel_progressarticle, efros2016origin, galland2011two, galland2012lifetime}, 
nanowires \cite{Pavel_progressarticle, Pavel_universality}, 
nanorods \cite{Pavel_universality}, 
organic semiconductors \cite{ruth2017fluorescence}, 
molecules \cite{bout1997discrete, dickson1997onoff, hasse2004exponential},
and other systems \cite{berhane2017photoinduced, 2014gammelmark043839}.
In these quantum emitters, the changes of state are typically caused by physical mechanisms that fluctuate randomly (as opposed to periodically) and often without regard for their history.
The observed fluorescence provides a convenient connection to the underlying state and it is important to understand observed blinking statistics in order to analyse the cause of the state switches.
Key insights can be distilled from the switching rates and how these depend on relevant parameters, and yet these rates are often the hardest to extract from the raw time series of photon counts in particular for low-intensity light signals or noisy data.

\begin{figure}
	\includegraphics[width=159.4mm]{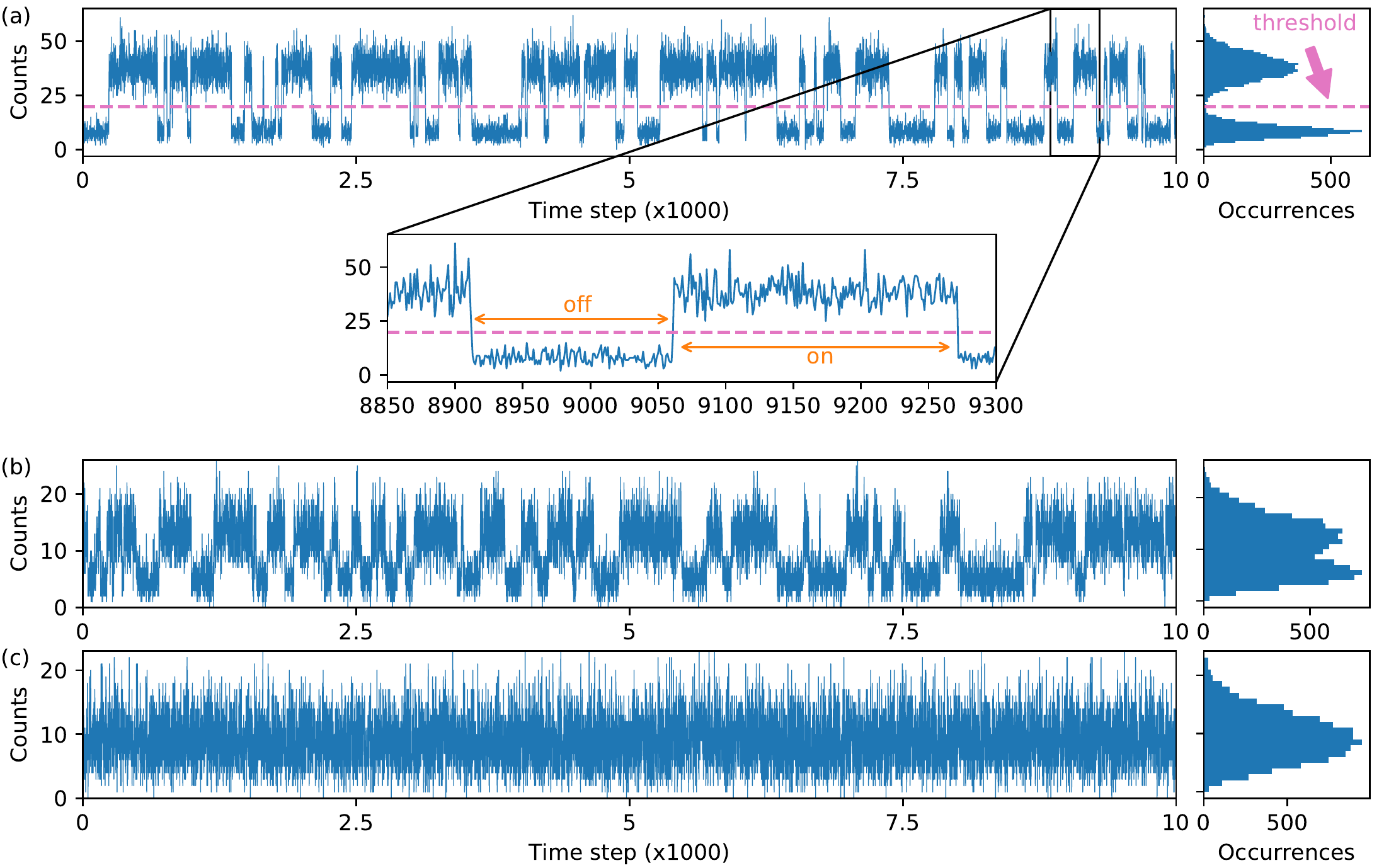}
\caption{
	Not all blinking time traces suit analysis in terms of a threshold.
	(a) Simulated data with clear \textit{on} and \textit{off} states.
	The dotted line represents a threshold, which is used to distinguish the \textit{on} and \textit{off} states.
	(b) Simulated blinking time trace with higher background noise but same switching rates as in (a).
    (c) A high rate of switching events relative to the detector interval makes it harder even to see the blinking
}
  \label{fig:PL}
\end{figure}

Blinking typically leads to step-like switches in the fluorescence time trace, as illustrated in Fig.~\ref{fig:PL}(a).
The most common method used to analyse the \textit{on} and \textit{off} states from a blinking time series is threshold analysis \cite{jantzen2016nanodiamonds, Kuno_nonexponential, Kuno_onoff}, illustrated for simulated data in Fig.~\ref{fig:PL}(a).
However, the choice of threshold intensity is essentially arbitrary, and it can significantly influence the statistics of the \textit{on} and \textit{off} states \cite{Catherine_facts, Pavel_model, watkins2005detection, amecke2014distortion}.
The threshold technique becomes difficult to use for emitters fluorescing at low signal-to-noise ratios as depicted in Fig.~\ref{fig:PL}(b), where the distributions of counts from the \textit{on} and \textit{off} states to overlap.
Thresholds are not at all helpful if the blinking switches at rates high relative to the detector interval as shown in an extreme case in Fig.~\ref{fig:PL}(c).
Here we present a method utilising Bayesian statistical inference to find the blinking rates directly from the photon-count data series, without requiring a threshold. 
The method works for blinking quantum light emitters independent of their brightness or speed of switching. 
This method is shown to be more accurate than threshold analysis, more broadly applicable, and more capable of assessing the measurement uncertainty.
Hence our method  allows for a better understanding of the mechanism behind the actual blinking phenomenon.

We model the blinking emitters on hidden, discrete and continuous-time Markov processes. 
The accumulated photon counts over a detector interval obscure or hide the underlying process which is assumed to be memoryless. 
This naturally models a system whose time spent in each \textit{on} and \textit{off} state decays exponentially \cite{efros1997random, jantzen2016nanodiamonds, berhane2017photoinduced}.
The Bayesian inference on this model then allows us to determine the underlying switching rates of these hidden states. 
Some more complex physical systems have been found to have \textit{on} and \textit{off} intervals instead distributed according to a power law 
\cite{Kuno_nonexponential, cichos2007power, tang2005diffusion, tang2005mechanisms},
indicating non-Markovian behaviour, or memory effects in the system.
This precludes the independencies that lead to elegant factorisation in the derivations presented here, and so the computational processes would need to be significantly extended to tackle such data.
However, the problems identified here for threshold analysis certainly also apply to non-exponential distributions.
In fact, reliable discrimination between exponential and power law dependence is only possible by careful and reliable analysis of the blinking data.

We have modelled three behaviours for the state relative to the detector interval.
In section \ref{sec:DTMC single interval} we analyse a simple blinking process using a discrete time Markov chain, where the emitter is assumed to be either \textit{on} or \textit{off} during each detector interval and switching of state can occur only at the interval boundaries.
The extreme case of continuous switching within the detector intervals is analysed in section \ref{sec:CTMC} using a continuous time Markov process.
In real experiments only a finite number of switching events are likely to occur within any given detector interval rather than no switching or arbitrary switching events.
Such a process is discussed in section \ref{sec:DTMC multi-step} using a discrete time Markov process and considering subintervals inside the detector interval.
Finally in section~\ref{sec:inferring_the_state} we show how to use the Bayesian analysis to infer the state at each time step.

\section{Discrete-time Markov Chain single-step Model}
\label{sec:DTMC single interval}

\subsection{Model description}

\begin{figure}
\centering
\includegraphics[width=0.3\columnwidth]{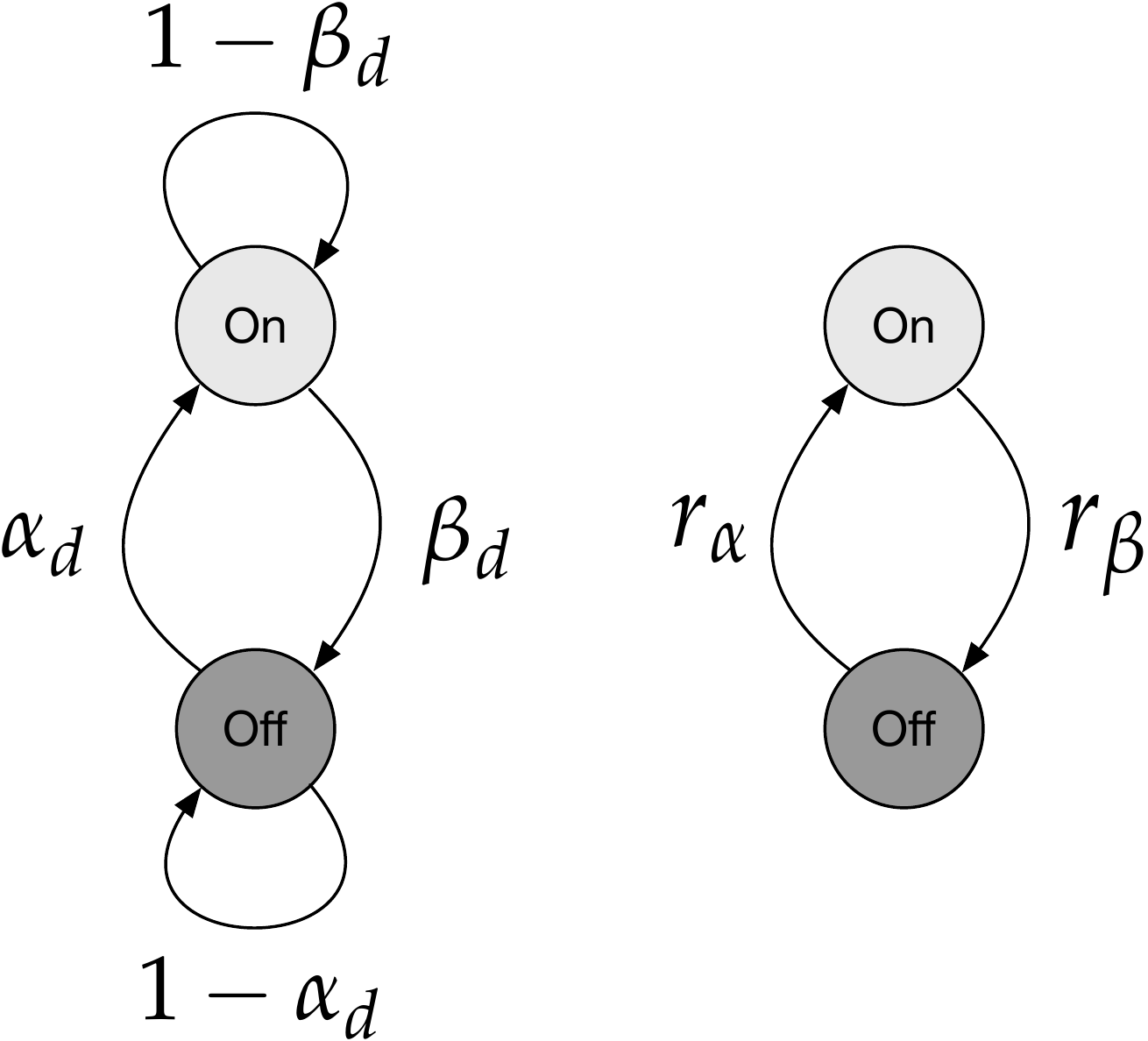}
\caption{
  Markov process modelling the evolution of the state. 
  Left: a discrete time Markov process with $d$ time steps per detection interval. 
  In this case $\dpon$ and $\dpoff$ are the switching probabilities. 
  Right: a continuous time Markov process where $\ron$ and $\roff$ are switching rates. 
  }
\label{fig:DTMC}
\end{figure}

The critical step in making an inference on the blinking rates from the observed fluorescence data is constructing a model of how the counts arise given the structure of the problem.
We consider that the emitter has only two relevant states that affect the fluorescence levels: an \textit{on} state and an \textit{off} state, and there is some mechanism that causes the emitter to switch between these states.
We model the switching as a Markov process, where the current state is sufficient to determine the future dynamics.
For simplicity, we begin by considering the state to be  fixed for the entire detector interval and it can only switch at the boundaries with switch-on and switch-off probabilities $\alpha_1$ and $\beta_1$ respectively, where the subscript represents the allowed number of state changes per detector interval.
This notation anticipates a discrete-time Markov chain (DTMC) which allows for multiple switching events per detector interval, which is discussed in later sections.
The Markov chain for this more general model is depicted on the left in Fig.~\ref{fig:DTMC}, where $d=1$ corresponds to the single switching opportunity case.

The blinking model also contains the fluorescence and background rates $\lambda$ and $\mu$.
If the emitter is in an \emph{off}-state then the detector can still register counts (known as dark-counts) due to noise processes in the detector or electronics. 
These counts are well modelled by a Poissonian process with a rate $\mu$. 
While fluorescing, it is assumed that the emitter is being driven by a strong classical pump so the statistics of the photon counts are also well modelled by a Poissonian process (i.e. shot noise) with an overall rate $\lambda$ that includes detector inefficiencies.
The \emph{observed} counts are then Poissonian with a rate of either $\mu$ or $\mu+\lambda$ since the combination of two independent Poissonian processes is again Poissonian. 
Given $\lambda$, $\mu$, and the state $s_{t-1}$ at the beginning of time-step $t$, the probability of seeing $c_t$ counts is 
\begin{equation}
    P(c_t|s_{t-1}\,\lambda\,\mu\, I_1) = \left\{ \begin{array}{cc}
       \poisson{\mu}{c_t}  &\quad s_{t-1}=0 \\
       \poisson{\mu+\lambda}{c_t}   &  \quad s_{t-1}=1
    \end{array} \right.
\label{eqn:poisson eqn}
\end{equation}
where $s_{t-1} = 1$ represents the \textit{on} state and $s_{t-1} = 0$ represents the \textit{off} state for interval $t$, $I_1$ tags the single interval model as the background information, and $\poisson{\ell}{c}=c^\ell \exp(-\ell)/c!$ is the probability of observing $c$ counts given a Poisson distribution with rate $\ell$.

Clearly, if we knew the state of the emitter at each data point it would be a simple matter to infer the switching rates. 
Unfortunately these states are not directly observable, and worst still, because the states are not observed the switching probabilities become dependent on the entire history of the count data. 
This makes the inference considerably more involved. 

\begin{figure}
    \begin{center}
		 \includegraphics[width=0.5\columnwidth]{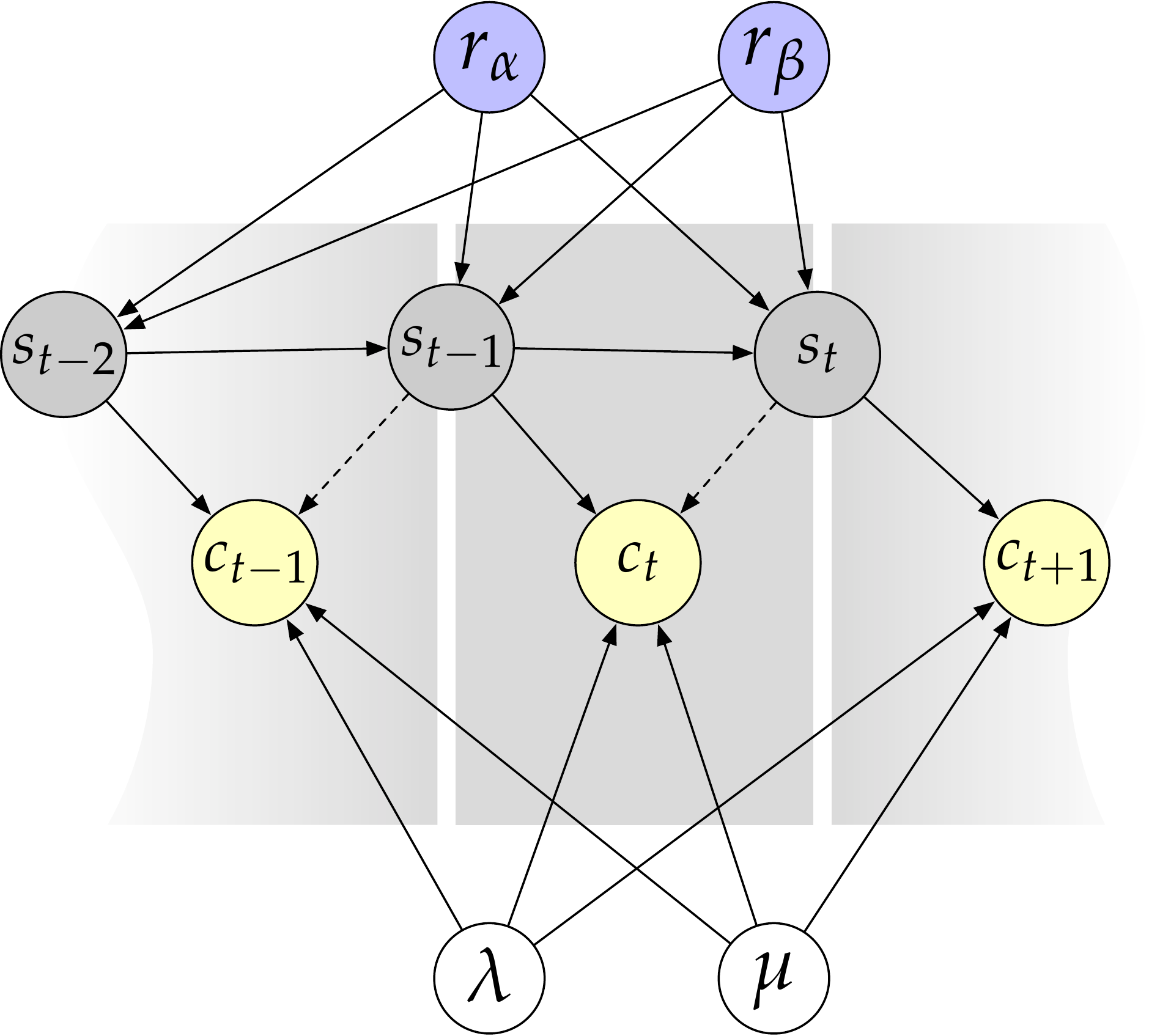}
    \end{center}
  \caption{
	  A partial Bayesian network representing the joint probability distribution of problem parameters. 
	  The pattern of nodes and connections inside the central grayed region representing detection interval $t$ are repeated for each data value. 
	  We are estimating parameters $\ron$, $\roff$, $\lambda$, and $\mu$ given the observed counts $c_t$. 
    For the discrete-time models $\ron$ and $\roff$ represent the probabilities $\dpon$ and $\dpoff$ respectively.
	  Enough of the full network is drawn to be able to easily determine the variable independencies. 
    In the case of a single step over the detector interval, $s_{t-1}$ is the state over the entire interval and the dashed arrows are absent.
    In all other cases $s_{t-1}$ and $s_{t}$ are the states at the boundaries of the detector interval.
	  }
\label{fig:BN}
\end{figure}

We can summarize the dependencies amongst the variables by the Bayesian network (BN)  \cite{darwiche2009modeling} shown in Fig.~\ref{fig:BN}.
The BN can be used to determine conditional independencies between the problem variables, using the property of $d$-separability.
In the BN, $s_{t-1}$ and $s_{t}$ represents the state of the emitter at the boundaries of the $t^\mathrm{th}$ detector interval, but for the single-step model $I_1$ the initial state extends for the whole duration so the dashed arrows in the figure should be omitted. 
That is, Eq.~\eqref{eqn:poisson eqn} depends only on $s_{t-1}$ (this will change for multi-step models).
The key variable independencies are the following
\begin{align}
    \pon \perp \poff \perp \lambda \perp \mu |I_1 \label{eq:I1indep-unobserved}\\
    s_t \perp s_{u} |s_{t-1}\,\Omega_1 \quad (u<t-1) \label{eq:I1indep-steps}\\
    c_t \perp c_u, s_v | s_{t-1}\,\Omega_1 \quad (u\ne t, v\ne t) \label{eq:I1indep-counts}
\end{align}
where $\Omega_1 = \pon\,\poff\,\lambda\,\mu\,I_1$ for compactness.
We will make use of these independencies in the following section.

\subsection{Bayesian Inference}\label{sec:single interval inference}

Ultimately, the quantity we want to infer is $P(\pon\,\poff|\vec{c}\,I_1)$ where $\vec{c}\equiv c_1\,c_2\ldots c_N$ is the set of count data obtained over $N$ detector intervals.
Using Bayes' rule, the posterior probability distribution of $\pon$ and $\poff$ can be written as,
\begin{equation}
P(\pon\,\poff|\vec{c}\, I_1) = \frac{P(\pon\, \poff| I) P(\vec{c}\,| \pon\, \poff\, I_1)}{P(\vec{c}\,|I_1)}
\end{equation}
The rates of fluorescence and background counts, $\lambda$ and $\mu$ can be added then removed by marginalisation,
\begin{align}
	P(\pon\, \poff|\vec{c}\, I_1) &= \int\!\! d\lambda\int\!\! d\mu \frac{P(\pon\, \poff| I_1) P(\vec{c}\,\lambda\,\mu| \pon\, \poff\, I_1)}{P(\vec{c}\,|I_1)} \\
	&= \frac{1}{P(\vec{c}\,|I_1)} \int\!\! d\lambda\int\!\! d\mu P(\pon \poff \lambda\, \mu| I_1) P(\vec{c}\,| \Omega_1). \label{eq:Pabgc}
\end{align}
Note that if $\mu$ and $\lambda$ are known independently then the integrals in \eqref{eq:Pabgc} are unnecessary and this will be true in all the models we develop.
Given \eqref{eq:I1indep-unobserved} we can factor $P(\pon\,\poff\,\lambda\,\mu|I_1)=P(\pon| I_1)P(\poff| I_1)P(\lambda| I_1)P(\mu| I_1)$.  
We will go further and take them all as constant over some initial range so that
\begin{align}
P(\pon\,\poff|\vec{c}\,I_1) &= \frac{1}{\mathcal{N}} \int\!\!d\lambda\int\!\!d\mu P(\vec{c}\,|\Omega_1)
\label{eqn:preinference}
\end{align}
and with the normalisation factor $\mathcal{N}$ to be determined at the end. 

As stated earlier, if all the states $\vec{s}\equiv s_0\,s_1\ldots s_{N-1}$ were known together with $\Omega_1$ the probability of all the counts would be easily determined. 
We can use the same trick of adding these parameters then marginalising over them so that,
\begin{align}
P(\vec{c}\,|\Omega_1) = \sum_{\vec{s}} P(\vec{c}\,\vec{s}\,|\Omega_1) = \sum_{\vec{s}} P(\vec{c}\,| \vec{s}\,\Omega_1) P(\vec{s}\,| \Omega_1)
\label{eqn:p(D)} 
\end{align}
where $\sum_{\vec{s}} = \sum_{s_0}  \ldots  \sum_{s_{N-1}}$. Using the independencies \eqref{eq:I1indep-counts}, $P(\vec{c}\,| \vec{s}\,\Omega_1)$ can be simplified to 
\begin{align}
    P(\vec{c}\,| \vec{s}\,\Omega_1) = \prod_{t=1}^{N}P(c_t|s_{t-1}\Omega_1),
\end{align}
and each term is determined by \eqref{eqn:poisson eqn}.

The remaining $P(\vec{s}\,| \Omega_1)$ term cannot be simply factorised over the states despite being a Markov chain, as observing $\Omega_1$ introduces possible dependencies. We can however expand using the product rule and simplify by making use of the independencies in \eqref{eq:I1indep-steps}:
\begin{align}
P(\vec{s}\,| \Omega_1) = \prod_{t=1}^{N}P(s_{t}|s_{t-1}\,\Omega_1)P(s_{0}|\Omega_1),
\end{align}
where we have chosen to expand in temporal order.

Finally, the inference becomes
\begin{align}
    P(\pon\,\poff|\vec{c}\, I_1) =& \frac{1}{\mathcal{N}} \int\!\!d\lambda\int\!\!d\mu \sum_{\vec{s}} \prod_{t=1}^{N} P(c_t\,s_{t}|s_{t-1}\,\Omega_1) \nonumber \\
    & \times P(s_0|\Omega_1) \label{eq:d1exponentalinference}
\end{align}
where we have used \eqref{eq:I1indep-counts} to write the joint distribution between $c_t$ and $s_t$.

The problem with \eqref{eq:d1exponentalinference} is that the sum over $\vec{s}$ contains $2^N$ terms each of which has $N$ products. 
This will rapidly become intractable as the size of the data grows. 
Fortunately it's possible to rewrite \eqref{eq:d1exponentalinference} as a \emph{single} term with $N$ $2\times 2$ matrix products. 
Consider the following matrix:
\begin{align}
    R_t = \begin{bmatrix}
      P(c_t\,s_{t}\!=\!0|s_{t-1}\!=\!0\,\Omega_1) & P(c_t\,s_{t}\!=\!0|s_{t-1}\!=\!1\,\Omega_1) \\
      P(c_t\,s_{t}\!=\!1|s_{t-1}\!=\!0\,\Omega_1) & P(c_t\,s_{t}\!=\!1|s_{t-1}\!=\!1\,\Omega_1)
  \end{bmatrix},\label{eq:Rmat}
\end{align}
and vector
\begin{align}
    \vec{D}_0 = \begin{bmatrix}
      P(s_{0}\!=\!0|\Omega_1) \\
      P(s_{0}\!=\!1|\Omega_1)
  \end{bmatrix}\label{eq:Dvec}
\end{align}
then 
\begin{align}\label{eq:matrixequiv}
    \sum_{\vec{s}} \prod_{t=1}^{N} P(c_t\,s_{t}|s_{t-1}\,\Omega_1)P(s_0|\Omega_1)=[1\; 1] \prod_{t=1}^N R_t \vec{D}_0
\end{align}
where the product on the right hand side is read as decreasing in $t$ to the right. 
The equivalence in \eqref{eq:matrixequiv} is readily verified by expanding a few terms out.
The matrix multiplication will sum over the columns of $R_t$ which is a summation over the states $s_{t-1}$. 
Each successive multiplication sums over another state and the final multiplication by the row vector $[1\;1]$ will sum over $s_N$.

With this equivalence, the inference reads
\begin{align}
    P(\pon\,\poff|\vec{c}\, I_1) =& \frac{1}{\mathcal{N}} \int\!\!d\lambda\int\!\!d\mu 
    [1\; 1] \prod_{t=1}^N R_t \vec{D}_0, \label{eq:singleintervalmatrix}
\end{align}
which is easily computable. 
Note that care must be taken to avoid underflow or overflow in calculating the matrix products.

\begin{figure}
	\includegraphics[width=159.4mm]{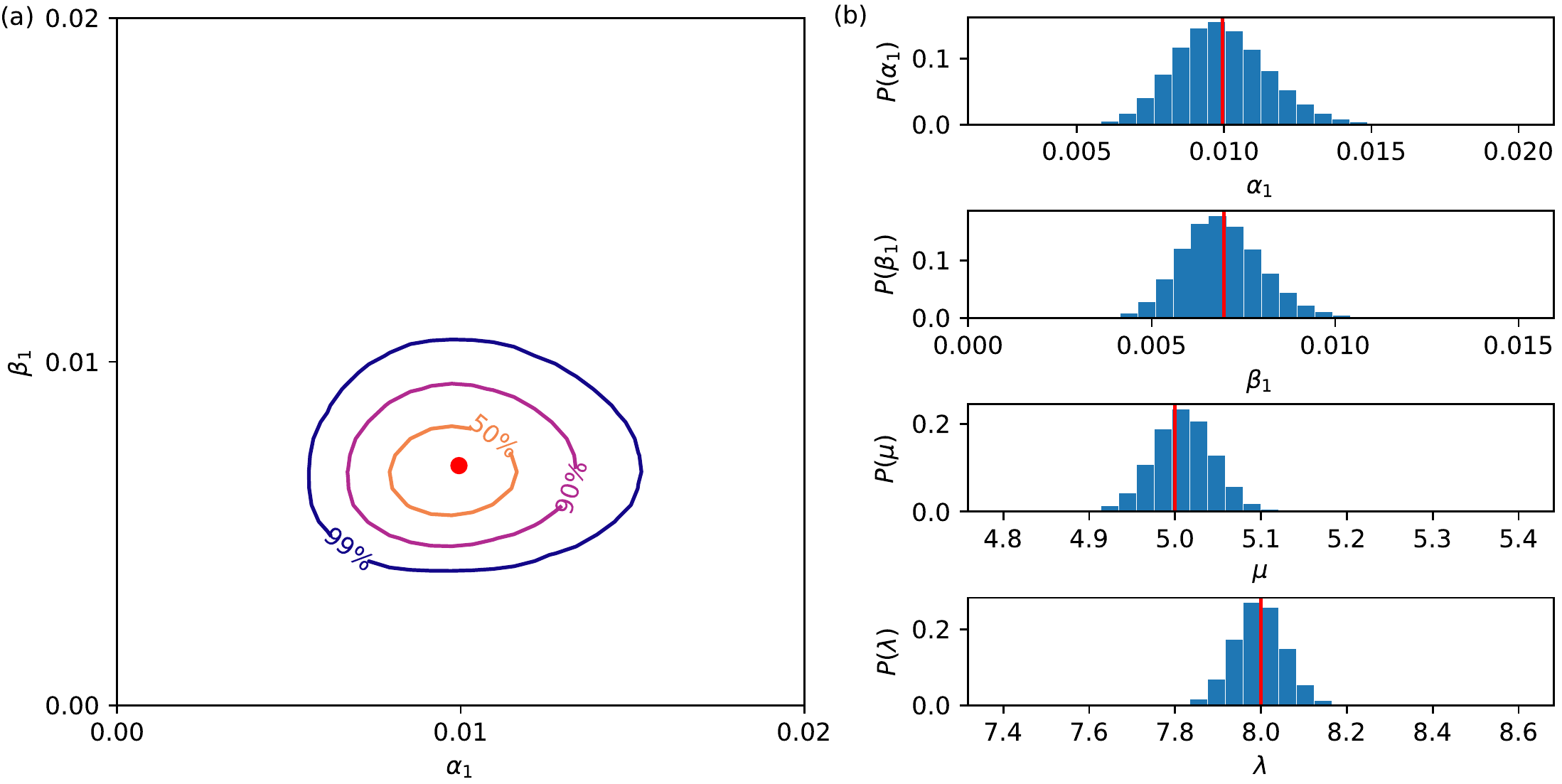}
	\caption{
		(a) Inference of the switching probabilities on simulated data given in Fig.~\ref{fig:PL}(b). 
	    The contours contain the 50\%, 90\%, and 99\% credible regions.
	     The true value is shown as red dot.
	    (b) Probability distributions of $\pon$ and $\poff$ and marginal distributions of $\mu$ and $\lambda$. The true values of the unknown parameters are given in red bar. 
    }
  \label{fig:single-interval_inference}
\end{figure}

\begin{figure}
	\hspace{24mm}\includegraphics[width=110mm]{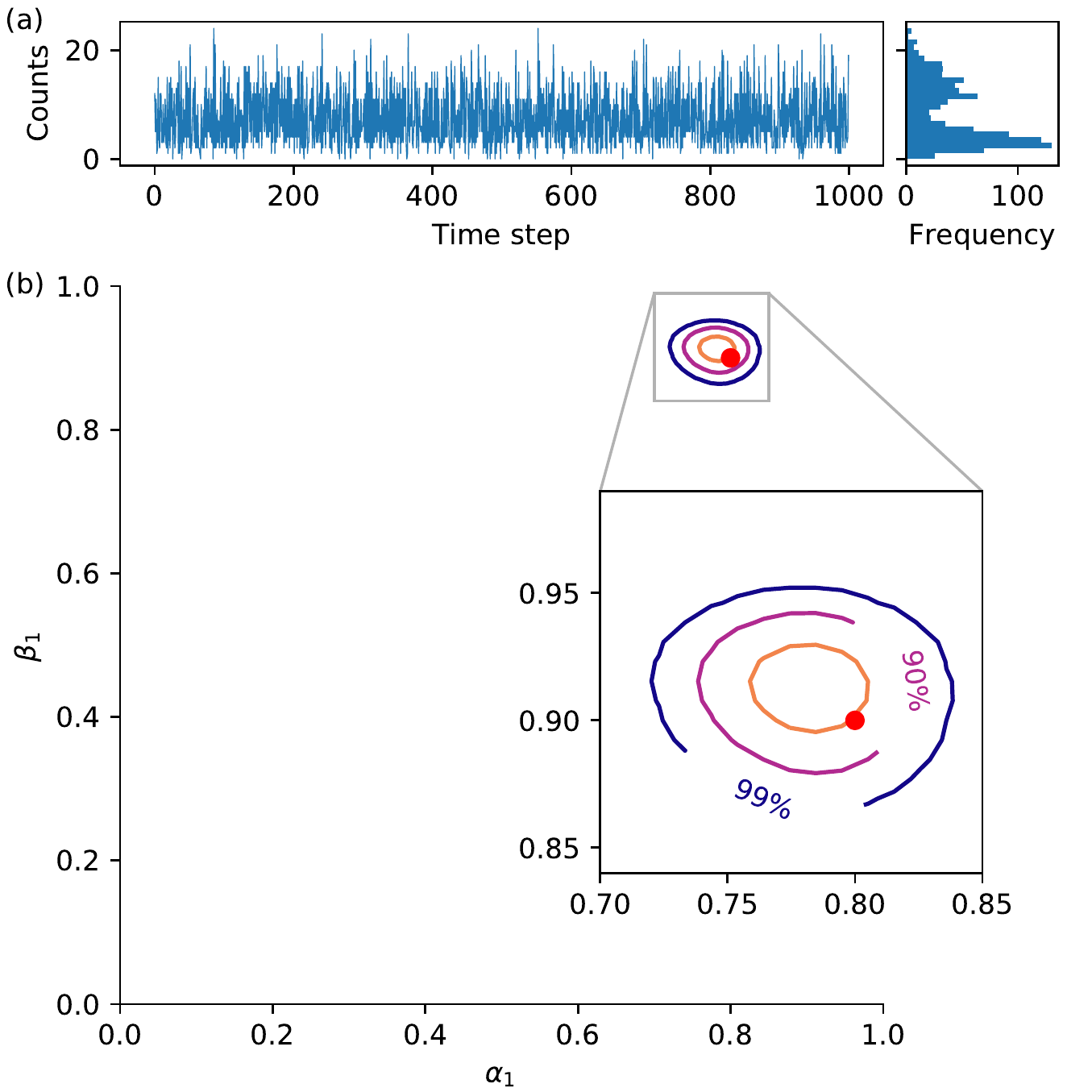}
	\caption{
		(a) Time trace with high switching probability of $\pon = 0.8$ and $\poff =0.9$, simulated using the DTMC single-step model. 
		(b) The posterior distribution for switching probabilities inferred using single interval model.
    }
  \label{fig:single-interval_inference_high switching rate}
\end{figure}

As a demonstration of the algorithm, Fig.~\ref{fig:single-interval_inference} (a) shows the credible regions of $\pon$ and $\poff$ for the data given in Fig.~\ref{fig:PL}(b). 
The marginal distributions of all unknown parameters ($\pon$, $\poff$, $\mu$ and $\lambda$) are given in Fig.~\ref{fig:single-interval_inference} (b).
It should be noted that the inference contains no approximations or arbitrary choices, and makes use of all the data. 
The key point of contention with observed data is whether the model is realistic to the system. 
If it is, then the inference faithfully converges on the underlying values regardless of what they are. 
For example in Fig.~\ref{fig:single-interval_inference_high switching rate} we generated count data assuming high probabilities of switching ($\pon = 0.8$ and $\poff = 0.9$) so the data looks very unlike the usual blinking trace. 
The inference quickly converges on the correct probabilities despite this. 

\subsection{Comparison with threshold analysis}\label{sec:thresh_comparison}

\begin{figure}
	\includegraphics[width=159.4mm]{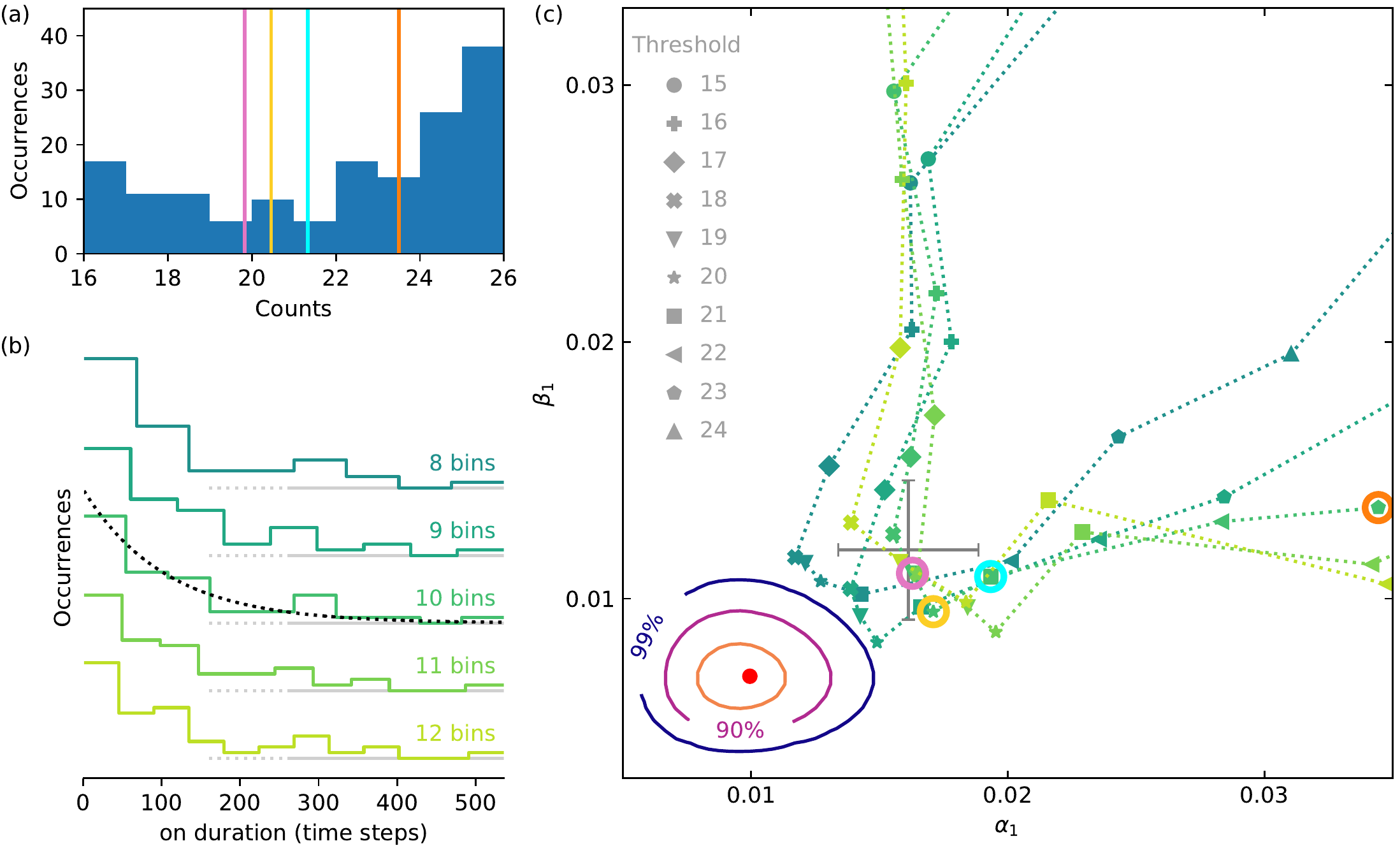}
	\caption{
		Comparison of Bayesian inference with threshold analysis.
		(a) Thresholds try to identify the states by a separation point between the two peaks in the histogram of counts.
		This is a close-up of the histogram in Fig.~\ref{fig:PL}(a), with four thresholds marked according to various rules found in the literature (see footnote).
		%
		%
		(b) A threshold allows the \textit{on} and \textit{off} periods to be identified, and the distribution of durations is obtained from histograms.
		Here the histogram of \textit{on} durations is shown with various (arbitrary) choices of binning.
		An exponentially decaying fit to this histogram (illustrated for 10 bins) yields the switching rate.
		(c) Switching probabilities for the data in Fig.~\ref{fig:PL}(a) were determined with thresholds ranging from 15 to 24 counts, and for the binning options illustrated in (b).
		Results for the four thresholds marked in (a) are circled in corresponding colours for 10 bins.
		The grey mark with error bars represents the mean and standard deviation of the range of points obtained for $17\leq I_\mathrm{threshold} \leq 21$, which are arbitrarily considered ``reasonable'' options.
		None of the arbitrary choices of threshold or binning yield a result lying within the 99\% credible region of the Bayesian inference.
	}
	\label{fig:threshold_comparison}
\end{figure}

In this subsection we compare our Bayesian inference technique with the conventional threshold analysis technique for extracting the \textit{on} and \textit{off} rates from a blinking time trace. 
To make threshold analysis possible, data is taken from the simulated count trace from Fig.~\ref{fig:PL}(a) which has well separated \textit{on} and \textit{off} states.
Given a threshold $I_\mathrm{threshold}$, the emitter is taken to be in the \textit{on} state if the intensity $I_n$ in the $n^\text{th}$ time bin satisfies $I_{n} > I_\mathrm{threshold}$ and it is \textit{off} if $I_{n} \leq I_\mathrm{threshold}$.  
A variety of methods have been employed for selecting the threshold, and four are considered here
\footnote{
	The four methods employed here are:
\begin{enumerate*}
\item
The minimum value between two peaks of the intensity histogram \cite{Catherine_facts}, which has been numerically determined here from a double-Poissonian fit.
This threshold is illustrated in Fig.~\ref{fig:PL}(a), and marked with the same magenta colour in Fig.~\ref{fig:threshold_comparison}(a).
\item
Two standard deviations above the mean intensity of background counts \cite{Kuno_nonexponential, Kuno_onoff, Catherine_facts, cordones2013mechanisms, knappenberger2007excitation}, indicated by a golden line in Fig.~\ref{fig:threshold_comparison}(a).
\item
The highest possible background count rate over total duration of the experiment \cite{Catherine_facts}.
Here we have taken this to be the position where the \textit{off} state Poissonian normalised to the number of measurement time steps dips below 1, and it is indicated by a cyan line in Fig.~\ref{fig:threshold_comparison}(a).
\item
The mid point between two peaks in the counts histogram \cite{Catherine_facts, cordones2013mechanisms}, marked with an orange line in Fig.~\ref{fig:threshold_comparison}(a).
\end{enumerate*}
}.
These are shown in Fig.~\ref{fig:threshold_comparison}(a) over a close-up of the centre of the counts histogram from Fig.~\ref{fig:PL}(a).

Once the \textit{on} and \textit{off} periods are identified they can be binned by duration into histograms \cite{efros1997random, jantzen2016nanodiamonds,  Kuno_onoff, Catherine_facts}.
Fig.~\ref{fig:threshold_comparison}(b) shows the \textit{on} duration histogram for a given threshold, highlighting the way that arbitrary choice of binwidth impacts the perceived distribution.
A separate histogram can be produced for the \textit{off} periods, and exponentially decaying fits to these histograms yield the characteristic \textit{on} and \textit{off} lifetimes from which the switching probabilities or rates are obtained \cite{efros1997random, jantzen2016nanodiamonds}.
This analysis was performed with a range of thresholds and of duration binwidths to explore how the outcomes depend on these choices, and the results are displayed in Fig.~\ref{fig:threshold_comparison}(c).
Results for the four marked thresholds are marked at the 10-duration-bins case because this was the default option in the package used for data processing (numpy).

Threshold analysis does not provide any mechanism for estimating the uncertainty in the extracted swithching rates.
This is because the fundamental measurement uncertainty arises from the difficulty in distinguishing short switching events noise spikes in the count time trace.
It is possible to acknowledge a kind of ``processing uncertainty'' by considering that various thresholds and bins are plausible, and accounting for the range of corresponding results.
The grey error-bar mark in Fig.~\ref{fig:threshold_comparison}(c) indicates the mean and standard deviation of all results obtained for $17\leq I_\mathrm{threshold} \leq 21$ and histograms with 8--12 bins.
The credible regions obtained from our Bayesian inference on the same blinking time trace is also shown in the figure. 
It is obvious from the picture that our Bayesian inference has converged close to the true value (the red dot in the Fig.~\ref{fig:threshold_comparison}(c)), and provided a meaningful assessment of the uncertainty.
None of the values obtained from threshold analysis are very close, and the Bayesian inference works better even for this time trace with very discernible blinking.

\section{Continuous-time Markov Chain Model}
\label{sec:CTMC}

In the previous section the emitter could only switch states at the boundaries of the detection interval.
Though this is a hidden Markov model, the inference was simple as the state is only obfuscated by the Poissonian distribution of counts.
In this section we go to the other extreme and assume the emitter can change state at any point in time, possibly multiple times in a single detection interval.
The previous single-interval model will be a good approximation to this continuous time model in the limit that the switching rates are very small.
In that limit the occasional switch during a detector interval introduces negligible error and most intervals are either entirely \emph{on} or entirely \emph{off}.

\subsection{Model description}
The emitter can switch at any time, so that the state is a function of time, i.e. $s(t)$. 
This situation is modelled by a continuous time Markov chain (CTMC) as depicted on the right of Fig.~\ref{fig:DTMC}. 
At any given time, the state can still only have one of two values but can switch arbitrarily often in any time interval. 
The detection events are as before with the detectors reporting the accumulated count from a time window of constant duration $T$.
For clarity we shall take time to be expressed in units of the detector interval so that $T=1$.

In a continuous-time Markov chain the time spent in any state is exponentially distributed, and since there are only two states, the next state is deterministic. 
The characteristic lifetime of the \textit{off}-intervals will be inversely proportional to the rate of switching \emph{out} of the \textit{off} state which will be the switch-on rate $\ron$.
Similarly the \textit{on}-intervals are inversely proportional to the switch-off rate $\roff$.
In this model, these two parameters $\ron$ and $\roff$, are the parameters we want to infer, replacing $\pon$ and $\poff$.

The CTMC can be solved by the forward Kolmogorov equations.
Representing the \emph{off} state by the vector $[1\; 0]^T$ and the \emph{on} state by $[0\; 1]^T$ the transitions dynamics are encoded by the equation
\begin{align}
    \frac{d P(t)}{dt} = Q P(t)
\end{align}
where
\begin{align}
    Q = \begin{bmatrix}
      -\ron & \ron \\
      \roff & -\roff
    \end{bmatrix}.
\end{align}
This has the formal solution $P(t) = \exp(Q t)$. 
From this solution we can determine the transition probabilities $P_{ab}(t)$ from state $a$ to state $b$ in time $t$:
\begin{align}
    P_{00}(t) &= \frac{1}{\ron+\roff}\left(\roff+\ron e^{-(\ron+\roff) t}\right), \\
    P_{01}(t) &= \frac{\ron}{\ron+\roff}\left(1-e^{-(\ron+\roff) t}\right), \\
    P_{10}(t) &= \frac{\roff}{\ron+\roff}\left(1-e^{-(\ron+\roff) t}\right), \\
    P_{11}(t) &= \frac{1}{\ron+\roff}\left(\ron+\roff e^{-(\ron+\roff) t}\right) .
\end{align}

\subsection{Bayesian Inference}

The basic inference is as before, we have a list of $N$ detector counts $\vec{c}$ and we want to infer $\ron$ and $\roff$ while optionally marginalising over the rates $\mu$ and $\lambda$ leading an equivalent expression to Eq.~\eqref{eqn:preinference} but for $\ron$ and $\roff$:
\begin{align}
P(\ron\,\roff|\vec{c}\,I_c) &= \frac{1}{\mathcal{N}} \int\!\!d\lambda\int\!\!d\mu P(\vec{c}\,|\Omega_c)
\label{eqn:preinference2}
\end{align}
where $\Omega_c = \ron\,\roff\,\lambda\,\mu\,I_c$ similar to previously, and the subscript denotes the continuous-time model, and $\mathcal{N}$ is suitably redefined.

To get traction on the problem, we add in the states at the \emph{boundaries} of the detection interval $s_j = s(t_j)$, where the $n^\mathrm{th}$ detection interval goes from $t_{n-1}$ to $t_n$ and $t_n=n$ since $T=1$.
\begin{align}
P(\vec{c}\,| \Omega_c) = \sum_{\vec{s}} P(\vec{c}\,| \vec{s}\,\Omega_c) P(\vec{s}\,| \Omega_c) 
\end{align}
Note that in this situation the states $s_j$ at the boundaries are not enough to specify the state during the detection interval, unlike in the previous case where the state was constant during the whole interval.
The indepencencies are now
\begin{align}
    c_n &\perp c_m | s_{n-1}\, s_n\, \Omega_c \quad  m\ne n \\
    s_n &\perp s_m | s_{n-1}\, \Omega_c \quad  m < n-1
\end{align}
and we can simplify
\begin{align}
    P(\vec{c}\,| \Omega_c) &= \sum_{\vec{s}} \prod_{n=1}^N P(c_n|s_{n-1}\, s_n\, \Omega_c)P(s_n|s_{n-1}\,\Omega_c)P(s_0|\Omega_c) \\
    &= \sum_{\vec{s}} \prod_{n=1}^N P(c_n\, s_n |s_{n-1}\,\Omega_c)P(s_0|\Omega_c).
\end{align}

This expression can again be converted into an efficient matrix equation identical in form to earlier:
\begin{align}\label{eq:ctmc-matrix}
    \sum_{\vec{s}} \prod_{n=1}^{N} P(c_t\,s_{n}|s_{n-1}\,\Omega_c)P(s_0|\Omega_c)=[1\; 1] \prod_{n=1}^N R_n \vec{D}_0
\end{align}
with
\begin{align}\label{eq:Rn-ctmc}
    R_n = \begin{bmatrix}
      P(c_n\,s_{n}\!=\!0|s_{n-1}\!=\!0\,\Omega_c) & P(c_n\,s_{n}\!=\!0|s_{n-1}\!=\!1\,\Omega_c) \\
      P(c_n\,s_{n}\!=\!1|s_{n-1}\!=\!0\,\Omega_c) & P(c_n\,s_{n}\!=\!1|s_{n-1}\!=\!1\,\Omega_c)
      \end{bmatrix},
\end{align}
and 
\begin{align}\label{eq:D0-ctmc}
    \vec{D}_0 = \begin{bmatrix}
      P(s_{0}\!=\!0|\Omega_c) \\
      P(s_{0}\!=\!1|\Omega_c)
    \end{bmatrix}
\end{align}

All that remains is to calculate $P(c_n\, s_n |s_{n-1}\,\Omega_c)$. 
If we knew $f$, the fraction of the interval for which the emitter was in the \emph{on} state, then the counts would be given by a weighted Poissonian.
That is, 
\begin{align}
    P(c_n\, s_n |s_{n-1}\,\Omega_c) = \int_0^1\!\! df P(c_n|s_n\, s_{n-1}\,f\,\Omega_c )P(s_n\, f|s_{n-1}\,\Omega_c),
\end{align}
and
\begin{align}
    P(c_n|s_n\, s_{n-1}\,f\,\Omega_c ) = \poisson{\mu+f \lambda}{c_n}. 
\end{align}

Now, we need to find an expression for $P(s_n\, f|s_{n-1}\,\Omega_c)$. 
For brevity, we'll introduce the notation $R_{ab}(f)\equiv P(s_n\!=\!b\, f|s_{n-1}\!=\!a\,\Omega_c)$.
This probability accounts for all the possible histories where the detection interval spent a fraction $f$ in state \emph{on} and ended in the state $s_n$ given it started in state $s_{n-1}$.
The different histories can be decomposed by differing numbers of switch events.
Consider the case where the state starts and ends in the \emph{off} state, i.e.  $R_{00}(f)$. 
Clearly an odd number of switch events will not be consistent with the boundary states and hence will have probability zero.
With zero switch events the entire interval is \emph{off} and the probability is just the exponential distribution $\delta(f)e^{-\ron}$, where $\delta$ is the Dirac delta function.
Two switch events partitions the interval into three regions $\leg_1$, $\leg_2$, and $\leg_3$ with states \emph{off}--\emph{on}--\emph{off} and probability
\begin{align}
    \int_0^{1-f}\!\!\!\! d\leg_1 e^{-\ron \leg_1}\ron e^{-\roff \leg_2}\roff e^{-\ron \leg_3} = \ron\roff e^{-\ron(1-f)-\roff f} (1-f)
\end{align}
where we have used $\leg_2=f$ and $\leg_1+\leg_3=1-f$.
In general the interval will be partitioned into a set of \emph{on} states of total duration $f$ and a set of \emph{off} states of duration $1-f$ so that the exponentials combine and will not depend on the durations $\leg_j$.
For example with six switches there are seven regions with three switch-on and three switch-off events, and the probability is
\begin{align}
    & \ron^3\roff^3 e^{-\ron(1-f)-\roff f} \nonumber\\
    &\times \underbrace{\int_0^f\!\!d\leg_2 \int_0^{f-\leg_2}\!\!d\leg_4}_A \nonumber\\
    &\times \underbrace{\int_0^{1-f}\!\!d\leg_1 \int_0^{1-f-\leg_1}\!\!d\leg_3 \int_0^{1-f-\leg_1-\leg_3}\!\!d\leg_5}_B\; 1 \label{eq:6switches}
\end{align}
The integrals in $A$ calculate the volume of a two-dimensional simplex of side $f$, and those in $B$ the volume of a three dimensional simplex of side $1-f$.
Since the volume of an $n$-dimensional simplex of side $f$ is $f^n/n!$, Eq.~\eqref{eq:6switches} evaluates to
\begin{align}
    \frac{f^2 (1-f)^3 \ron^3\roff^3 e^{-\ron(1-f)-\roff f}}{2!3!}.
\end{align}
Alternatively, these integrals can be evaluated using a Laplace transform as shown by Jaynes \cite{2003JaynesLogicScienceChap18}. 

Summing over all switch events gives
\begin{align}
    R_{00}(f) =& 
    \delta(f)e^{-\ron}\nonumber \\ &+ 
    e^{-\ron(1-f)-\roff f}\sum_{k=0}^\infty \frac{f^{k-1}(1-f)^k}{(k-1)!k!}\ron^k\roff^k \\
    =& \delta(f)e^{-\ron} - e^{-\ron(1-f)-\roff f}\sqrt{\frac{(f-1)\ron\roff}{f}}\nonumber \\
    & \times J_1\left(2\sqrt{(f-1)f\ron\roff}\right)
\end{align}
where $J_1(x)$ is the Bessel function of the first kind.
A similar argument yields the other possible fraction probabilities:
\begin{align}
  R_{01}(f) =& \ron J_0\left(2\sqrt{(f-1)f\ron\roff}\right) e^{-\ron(1-f)-\roff f} \\
  R_{10}(f) =& \roff J_0\left(2\sqrt{(f-1)f\ron\roff}\right) e^{-\ron(1-f)-\roff f} \\
  R_{11}(f) =& \delta(1-f)e^{-\roff} + e^{-\ron(1-f)-\roff f}\sqrt{\frac{f\ron\roff}{f-1}} \nonumber \\
  & \times J_1\left(2\sqrt{(f-1)f\ron\roff}\right).
\end{align}
So in summary we have
\begin{align}
\label{eqn:final_count_ctmc}
    P(c_n\, s_n |s_{n-1}\,\Omega_c) = \int_0^1\!\! df \poisson{\mu+f \lambda}{c_n} R_{s_{n-1}s_n}(f)
\end{align}
which is an integral that needs to be done numerically. 

\begin{figure}
	\includegraphics[width=159.4mm]{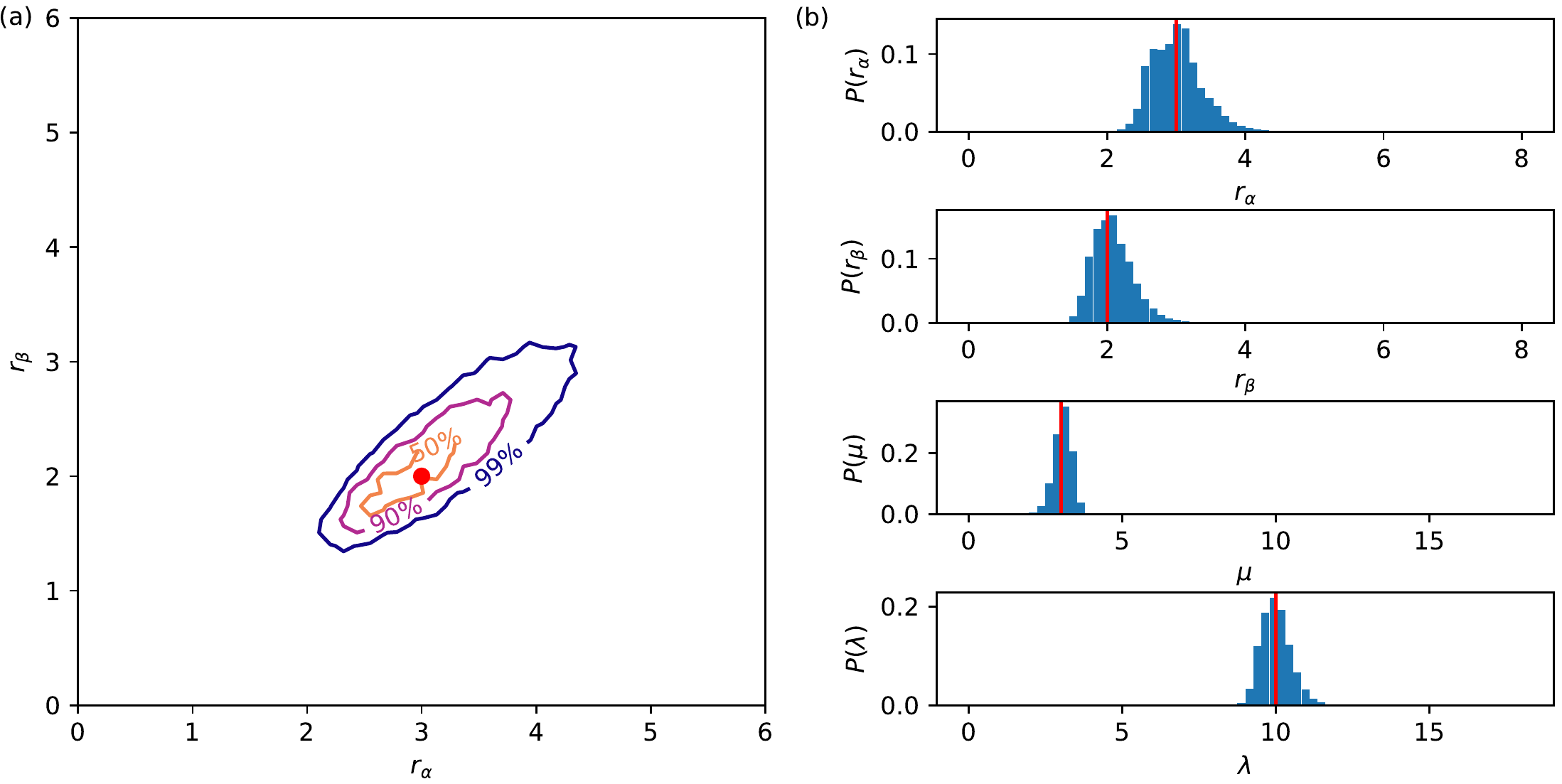}
	\caption{
	    (a) Inference of the switching rates on simulated data given in Fig.~\ref{fig:PL}(c). 
	    The contours contain the 50\%, 90\%, and 99\% credible regions.
	     The true value is shown as red dot.
	    (b) Probability distributions of $\ron$ and $\roff$ and marginal distributions of $\mu$ and $\lambda$. The true values of the unknown parameters are given in red bar. 
    }
  \label{fig:ctmc-interval_inference}
\end{figure}

The extreme case data in Fig.~\ref{fig:PL}(c) was used as an example of the inference and is presented in Fig.~\ref{fig:ctmc-interval_inference}. The initial range of parameters was assumed to be equally likely anywhere in $0\le\ron,\roff\le8$, and $0\le\lambda,\mu\le18$, and a 4-dimensional grid where each range was divided into 70 values was used in evaluating the posterior probability in \eqref{eq:ctmc-matrix} for each data point to allow for temporary normalization and control of underflow.

\subsection{Relation between single interval model and continuous-time model}

The single interval model and the continuous time model make different assumptions about the underlying switching mechanism, nevertheless if the switching rates $\ron$ and $\roff$ are low, we would expect the single interval model to yield very similar results. 
This is because the detector intervals that contain internal switching events will be relatively rare and will not contribute significantly to the inference.
Additionally, the discrete time model assumes that the emitter remains in a constant state over each detector interval, which is a reasonable approximation for small switching rates.
It might be supposed that, if the rates are small, the single interval model could be used as a more computationally efficient approximation to the full inference.
We now explore the validity of this approximation.

The two models can be connected by the time spent in a given state. 
Consider a time interval $\Delta t = T/d$ which is a fraction of the measurement time. 
In the continuous-time model the probability of a state having a lifetime $\Delta t$ is exponentially distributed, e.g $P_0(\Delta t)\equiv P(s(t)=0:0\le t \le \Delta t) = \exp(-\ron\Delta t)$. 
So for a model where the state is fixed for duration $\Delta t$, we can relate the probability of switching state to the probability of the state not having lifetime $\Delta t$, e.g. $\alpha_{d} = 1-P_0(\Delta t)$. 
Consequently,
\begin{align}\label{eqn:relation showing rates and probability}
    \dpon &= 1-e^{-\ron T/d}\\
    \dpoff &= 1-e^{-\roff T/d}
\label{eqn:relation showing rates and probability2}
\end{align}

Assuming small switching rates in the continuous model we can expand to first order in the switching rates. 
For Bessel function of the first kind,
\begin{equation}
J_m(z) \approx \frac{1}{\Gamma (m+1)}\left(\frac{z}{2}\right)^{m} \qquad z\ll 1
\end{equation}
where $\Gamma(x)$ is the Gamma function. Hence at a low rate of switching, we can expand first order in $\ron$, $\roff$ to obtain $R_{00}(f) \approx \delta(f)(1-\ron)$, $R_{11}(f) \approx \delta(1-f)(1-\roff)$, $R_{01}(f) \approx \ron$ and $R_{10}(f) \approx \roff$. 
Furthermore, from \eqref{eqn:relation showing rates and probability} and \eqref{eqn:relation showing rates and probability2} with $d=1$, $\pon \approx \ron$ and $\poff \approx \roff$. 
With this approximations \eqref{eqn:final_count_ctmc} yields,
\begin{align}\label{eqn_lowlimit_rate00}
P(c_n\, 0 |0\,\Omega_c) &\approx  (1-\pon)\,\poisson{\mu}{c_n}\\
P(c_n\, 1 |1\,\Omega_c) &\approx  (1-\poff)\,\poisson{\mu+\lambda}{c_n}\\\label{eqn_lowlimit_rate11}
P(c_n\, 1 |0\,\Omega_c) &\approx  \pon \tilde{c}_n\\
P(c_n\, 0 |1\,\Omega_c) &\approx  \poff \tilde{c}_n
\end{align}
where,
\begin{equation}
\tilde{c}_n = \int_0^1\!\! df \poisson{\mu+f \lambda}{c_n},
\end{equation}
is the probability of obtaining $c_n$ counts averaged over all durations spent \textit{on} in the interval.
The first two probabilities are identical to those obtained in the single interval model. 
The last two differ, as in the single interval model we made the arbitrary decision to take the switch event as happening at the ends of the intervals, whereas in the correspondence between models we allow the switch to occur anywhere. 
However, detector intervals that contain a switch event become rare when the rates are small. 
So the continuous time model becomes indistinguishable from the single step model in this limit as expected.

\begin{figure*}
	\centering{
		\includegraphics[width=\textwidth]{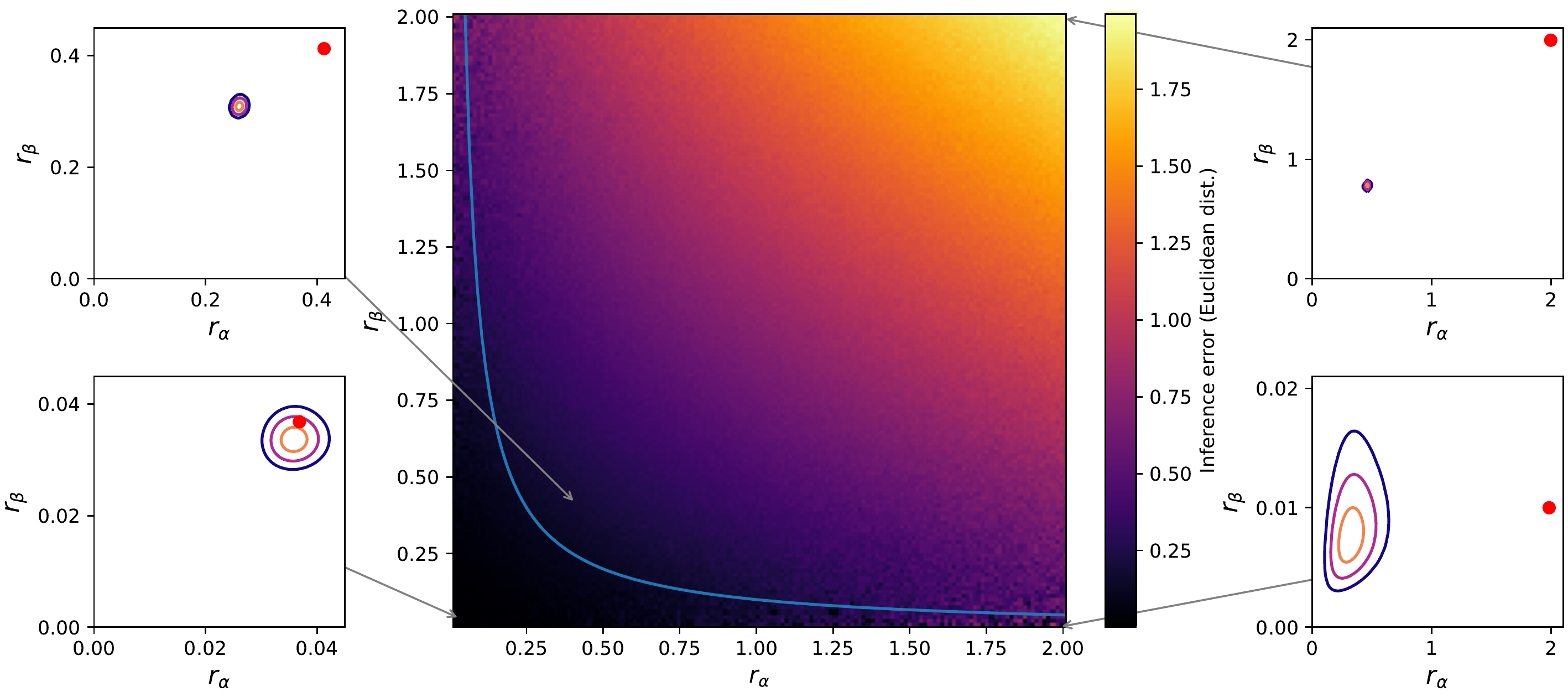}
	}
	\caption{
		Accuracy of the single-step inference to data simulated using a continuous model with various switching rates. 
		The inference error taken as the Euclidean distance between the true $\ron, \roff$ and the mode of the posterior distribution.
		For small switching probabilities the inference gives good agreement with the known parameters, but for higher switching probabilities the inference is less accurate. 
		The cyan line indicates an approximate threshold for applicability of the single-step model of $\ron\roff < 0.1$, which corresponds approximately to a relative error $\le 25\%$.
	}
	\label{fig:single_step_applicability}
\end{figure*}

The single interval model will lose accuracy as the rate of switching increases and this can be explored numerically.
In Fig.~\ref{fig:single_step_applicability} we summarise such an investigation by plotting the difference between the true switching rate (for data simulated from a continuous-switching model) and the result of a single-interval model inference. 
The colour map represents the Euclidean distance from the true value to the maximum value of the posterior probability distribution.
The black region corresponds to a small distance, and is therefore the most applicable region for the single-interval inference. 
As the (continuous) rates increase the single-interval model starts failing. 
An approximate threshold for applicability of the single-step model is when $\ron\roff < 0.1$, which corresponds roughly to a relative error $\le 25\%$. 
The continuous time model performs well throughout this entire region (and beyond) as shown in Fig.~\ref{fig:ctmc-interval_inference}, even though the traditional threshold technique cannot even be applied to the data, as illustrated in Fig.~\ref{fig:PL}(c).

A grid of $150\times 150$ simulations were run to produce the colour-map in Fig.~\ref{fig:single_step_applicability}, and the posterior distributions for switching rates are shown for four illustrative cases.
Note that the scales are different for each contour plot. 
The lower left plot shows a good convergence of the inference to the true value, where the rates are small. 
As both rates increase up the diagonal of the plane, the plots on the upper left and upper right show the increasing deviation of the inference from the true value. 
In the top right the switching rates are so high that there is an average of two switching events per detector interval. 
The posterior distribution is tightly converged since there are a large number of switching events, but it is significantly diverged from the true value. 
The lower right plot considers a situation where the \textit{on} rate is very high and \textit{off} rate is low. 
The convergence is not good for this case, since the asymmetry of switching rates results in very few switching off events---the data remains in the \textit{on} state for most of the time. 
The single-interval inference provides a good estimate for the slow rate, but dramatically underestimates the fast rate which is beyond what is assumed in the model.

\section{DTMC Multi-step Model}
\label{sec:DTMC multi-step}

In the CTMC model of the previous section the emitter may switch state at any time, and we derived a final integral over the distribution of counts that has to be evaluated numerically.
In general this may be difficult or inefficient, and in such cases the continuous model may be approximated by a discrete switching model that provides multiple possible switching points within each detector interval.

\subsection{Recursion relation}

Consider the $n^{\mathrm{th}}$ detector interval, having boundary states $s_{n-1}$ and $s_n$; and observed counts $c_n$.
We have already noted that this distribution is independent of the photon counts in adjacent detector intervals, and so the inference developed within this interval will be equally valid for all intervals in the measurement.
Furthermore, this holds \emph{within} a detector interval also; the count distribution in a given subinterval is independent of adjacent subintervals given knowledge of the states at the boundaries of the sub-intervals.

We introduce the notation $\sd_x$ to denote the state of the emitter at fraction $x$ within the detector interval ($0\le x \le 1$), where $\sd_0 = s_{n-1}$ and $\sd_1 = s_n$.
Also the counts in a fractional interval from time $u$ to time $v$ will be denoted $c_{u,v}$.
In particular, we introduce the mid-point state $\sd_{1/2}$ and then marginalize over it in order to identify a useful recurrence relationship for the photon count distribution.
This intermediate state breaks the original interval into two subintervals, and clearly there are only $c_n+1$ possible assignments for the subintervals $(c_{0,\frac{1}{2}},c_{\frac{1}{2},1})$ to give the total counts: $\{(c_n,0),(c_n-1,1),\ldots,(0,c_n)\}$.

In the following discussion frequent distinction will be made between the four sub-cases of the photon count distribution corresponding with the possible start and finish state possibilities $\sd_u$ and $\sd_{u+t}$ of an interval of length $t$. 
To make this clearer we introduce the following notation:
\begin{equation}
	f_{i,j}(t,\kappa)\equiv P \left( c_{u,u+t}\!=\!\kappa, \, \sd_{u+t}\!=\!j | \sd_u\!=\!i, \Omega_t \right)
\end{equation}
where $\Omega_t = \alpha_t \beta_t \lambda \mu I_t$ assuming that switching probabilities $\alpha_t$ and $\beta_t$ can be defined over the time $t$.
Using the introduced intermediate state $\sd_{1/2}$ gives rise to the expression
\begin{align}
	f_{i,j}(t,\kappa) =  &\sum_{c_d=0}^{\kappa} f_{i,0}(t/2,c_d) f_{0,j}(t/2,\kappa-c_d) \nonumber
	\\+& \sum_{c_d=0}^{\kappa} f_{i,1}(t/2,c_d) f_{1,j}(t/2,\kappa-c_d) \label{eq:recurrance}
\end{align}
which is the sum of the two {\em discrete convolutions} of count probability distributions corresponding to the two possible mid-interval states. 
This offers a fundamental recursion relation between the probability distributions of counts over successive halvings of the interval.

For a given interval $t$, $f_{i,j}(t,\cdot)$ is function of counts (a slice through a joint probability distribution).
Since in any given experiment there will be a maximum count in an interval, the function $f_{i,j}(t,\cdot)$ can be recorded as a single dimensional array.
In light of this, Eq.~\eqref{eq:recurrance} shows that values of the functions $f_{i,j}(t,\cdot)$ can be calculated numerically from knowledge of the function values for half the interval, $f_{i,j}(t/2,\cdot)$.

This recursion will apply in any system where the count distribution over a single time interval is independent of adjacent intervals given knowledge of the boundary states; 
in cases where a closed form for the count distribution is intractable this can be exploited to estimate the fully continuous distribution at low computing time cost.
As a demonstration of this process we now show that repeated application of Eq.~\eqref{eq:recurrance} facilitates fast and accurate estimates of the functions $f_{i,j}$ arising from the CTMC model.

\subsection{Model description}

The CTMC photon counts distribution has been shown to limit to Poissonian probabilities in the case when the switching probabilities are small, and the state of the emitter is constant for most measurement windows.  
Even when switching probabilities are high, however, the emitter state will be (nearly always) constant over time windows sufficiently small. 
We construct a model in which the detector interval is subdivided into multiple subintervals, and the emitter is allowed to switch state only at the boundaries of these subintervals (similar to the discrete model in section \ref{sec:DTMC single interval}).
To leverage the intrinsic ``halving'' suggested by the recursion relation discussed above, suppose then that a number of subintervals $d=2^m$ is chosen such that $m$ is integer.
If $1/d$ is a short enough duration that the functions $f_{i,j}(1/d,\cdot)$ are well-approximated by weighted Poissonians, then it is expected that this model will give close agreement with the CTMC inference.
For example, the region of applicability identified in Fig.~\ref{fig:single_step_applicability} suggests that $m$ should be chosen at least large enough so that $r_\alpha r_\beta < 0.1\times 2^{2m}$ for every $r_\alpha$ and $r_\beta$ under consideration.  
Since this bound increases exponentially with $m$, large switching rates may be accommodated by modest increases in $m$.

In this model the total photon count observed in a detector interval is the sum of all counts that arise in each subinterval within the detector interval.
This kind of system which switches its state within the detector interval can also modelled by DTMC.
The switching probabilities over a subinterval are $\dpon$ (probability of switching \textit{on}) and $\dpoff$ (probability of switching \textit{off}) as shown in Fig.~\ref{fig:DTMC}.
Here we want to infer the rates of switch-on ($\ron$) and switch-off ($\roff$) per detector interval.
The switch-on and switch-off probabilities $\dpon$ and $\dpoff$ can be written in terms of $\ron$ and $\roff$ as given in Eqs.~\eqref{eqn:relation showing rates and probability} and \eqref{eqn:relation showing rates and probability2}.

Using (as before) $\mu$ and $\lambda$ to denote expected background and signal photon counts over a unit interval, and $\dpon$ and $\dpoff$ to denote switching probabilities over a period of $1/d$, we have
\begin{align*}
f_{i,j}(1/d,\kappa)\approx \left\{ \begin{array}{cc}
       (1-\alpha_{d})\poisson{\mu/d}{\kappa}    &\quad i=0,j=0 \\
       \alpha_{d} \poisson{\mu/d}{\kappa}   &\quad i=0,j=1 \\
     \beta_{d}\poisson{(\mu+\lambda)/d}{\kappa}    &\quad i=1,j=0 \\
      (1- \beta_{d}) \poisson{(\mu+\lambda)/d}{\kappa}   &\quad i=1,j=1 \\
    \end{array} \right.
\end{align*}
The four cases of $f_{i,j}(1/d,\cdot)$ may be stored as arrays indexed by the counts  $\kappa$.
Equation~\eqref{eq:recurrance} may then be applied $m$ times to numerically approximate the values of $f_{i,j}(1,\cdot)$.

\subsection{Bayesian Inference}

\begin{figure}
	\hspace{24mm}\includegraphics[width=110mm]{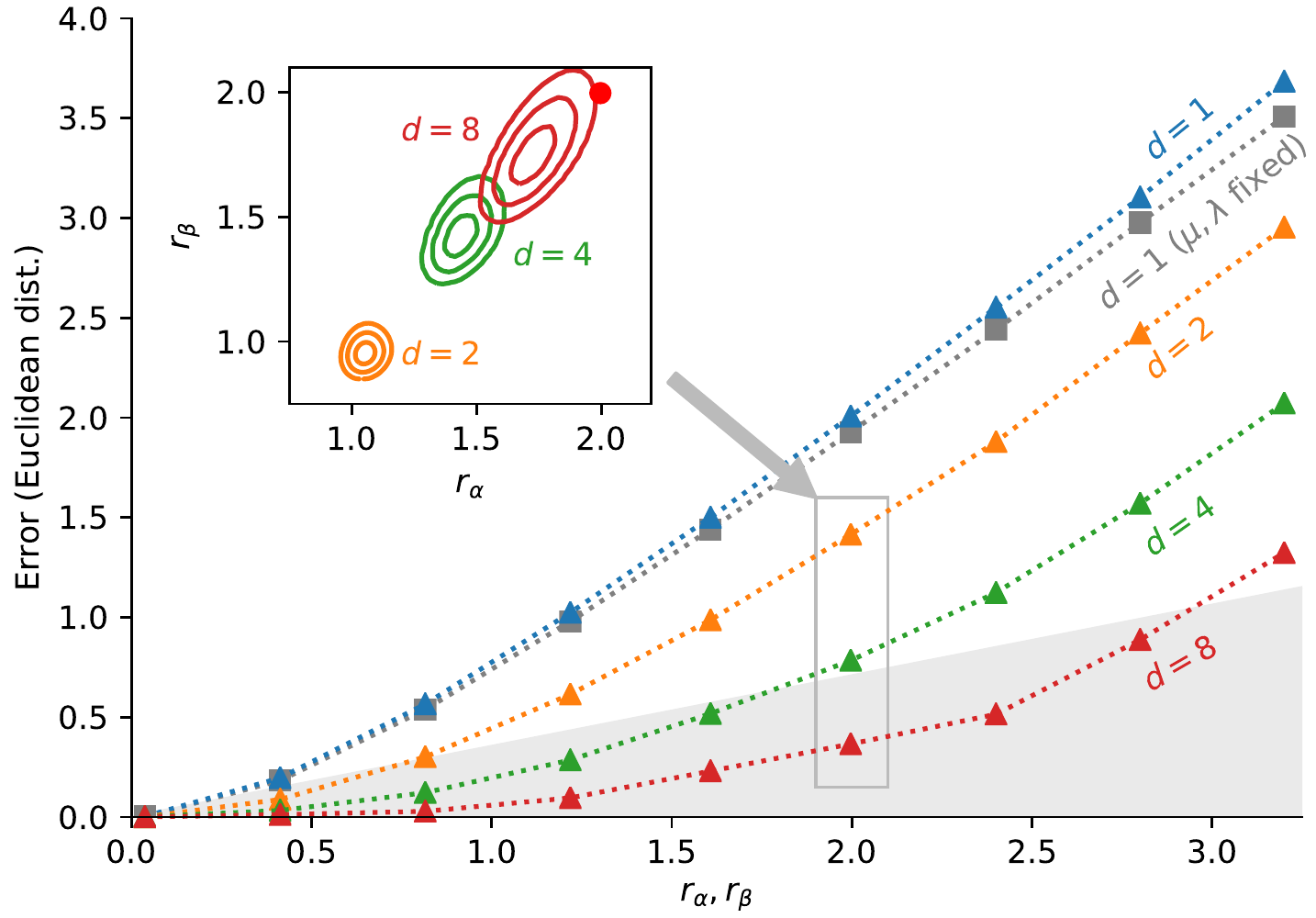}
	\caption{
		Increasing the number of subintervals considered in the inference reduces the error of the obtained switching rates.
		Simulated data with $r_\alpha = r_\beta$ were taken from the set generated for Fig.~\ref{fig:single_step_applicability} up to $r_\alpha = r_\beta = 2$ and additional datasets were generated for even faster rates.
		The multi-step inference was performed for various numbers of subintervals $d$ and the euclidean error determined by comparison of the known rates with the mode of the posterior distributions.
		The inset shows the 50\%, 90\%, and 99\% credible contours for the posterior distributions corresponding to $d=2, 4, 8$ for $r_\alpha = r_\beta = 1.9966$ (the contours for $d=1$ for this dataset are shown in Fig.~\ref{fig:single_step_applicability}).
		The shaded region represents a relative error of 25\%.
	}
	\label{fig:subinterval-advantage}
\end{figure}

As in the continuous time case, we are inferring the rates of switch-on $\ron$ and switch-off $\roff$ given the data counts $\vec{c}$. 
The inference proceeds almost identically, adding the unknown fluorescence $\lambda$ and background $\mu$ rates to the problem, using the independencies and finally  converting into a matrix form equivalent to the one given in \eqref{eq:matrixequiv}, 
\begin{align}\label{eq:matrixequiv_Id}
P(\vec{c}\,|\, \Omega_d) = [1\; 1] \prod_{n=1}^N R_n \vec{D}_0
\end{align}
with $R_n$ and $\vec{D}_0$ equivalent to Eqs.~\eqref{eq:Rn-ctmc} and \eqref{eq:D0-ctmc} respectively. 
Now we can use the recursion algorithm outlined to calculate $P(c_n\,s_{n}\,|\,s_{n-1}\,\Omega_d)$.

\begin{figure}
	\includegraphics[width=159.4mm]{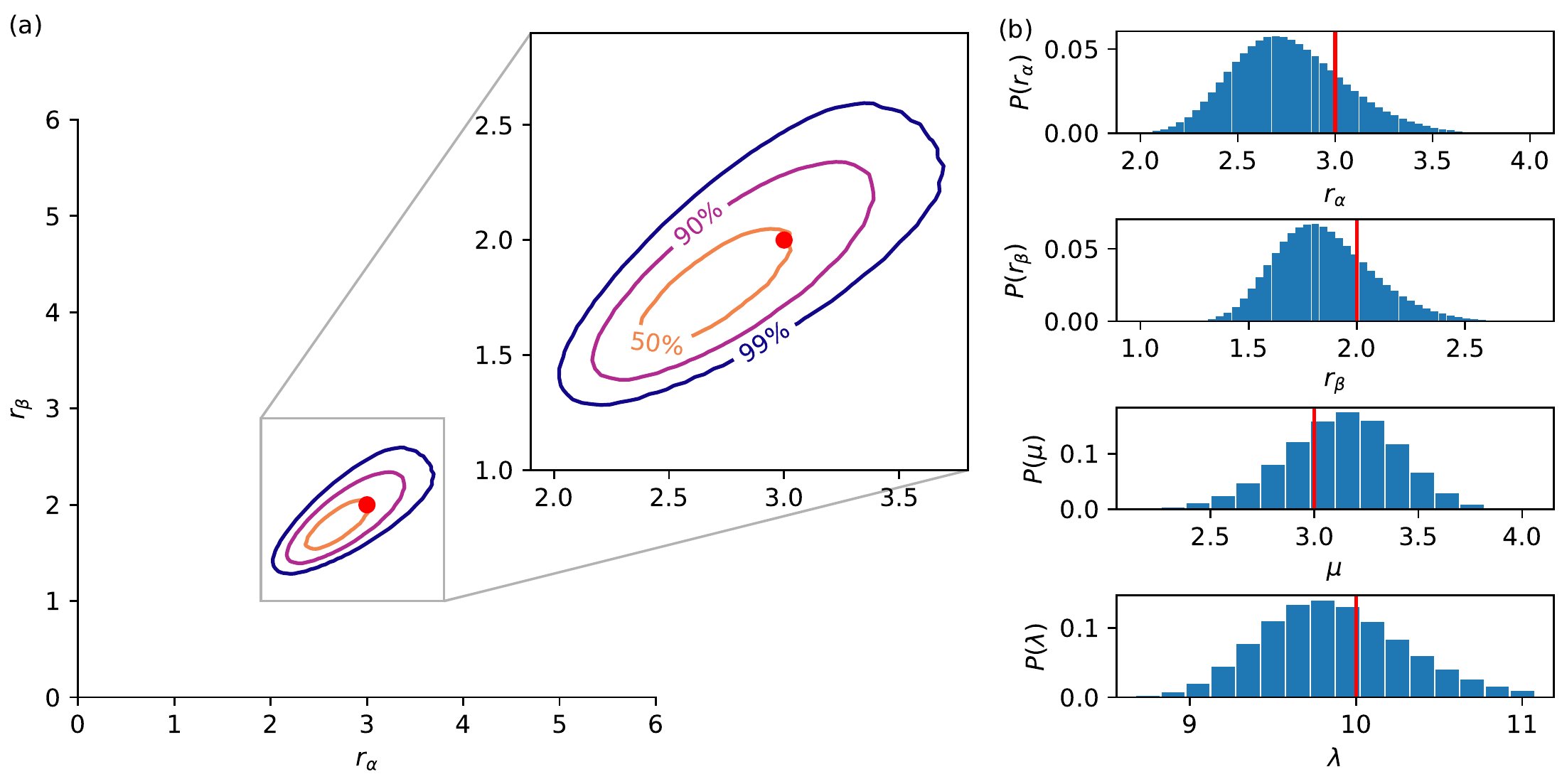}
	\caption{
		(a) Inference of the switching rates on simulated data given in Fig.~\ref{fig:PL}(c)
		DTMC multi-step model with $d=16$ subintervals was used for inference. 
		The contours contain the 50\%, 90\%, and 99\% credible regions.
		The true value is shown as red dot. 
		(b) Probability distributions of $\ron$ and $\roff$ and marginal distributions of $\mu$ and $\lambda$. The true values of the unknown parameters are given in red bar. 
    }
 \label{fig:dtmc-interval_inference}
\end{figure}

To demonstrate this inference and explore the significance of the number of subintervals $d$, we tackle set of data from the diagonal ($\ron = \roff$) of Fig.~\ref{fig:single_step_applicability}.
Three additional datasets beyond $\ron=\roff=2$ were generated, and inferences were calculated for $d=1, 2, 4, 8$.
Euclidean distance was again used to measure the error in the inference, and Fig.~\ref{fig:subinterval-advantage} shows the error increasing as $\ron, \roff$ increase.
Taking higher numbers $d$ of subintervals into consideration reduces the error, and $d=4$ is almost sufficient to keep the entire $\ron,\roff \le 2$ region from Fig.~\ref{fig:single_step_applicability} enclosed inside the 25\% relative error threshold.
The inset illustrates the posterior distributions that correspond to the third case in Fig.~\ref{fig:single_step_applicability}, except the plot scale is different.
As $d$ increases the location of the inference improves, but the width of the distribution converges less tightly.
This reflects the fact that for increasing $d$ there are higher numbers of intermediate states $\sd_j$ considered by the model but the amount of information provided by the dataset is not increasing.
In Fig.~\ref{fig:subinterval-advantage} each $d$ eventually leads to the error increasing with about the same slope, which is $\sqrt{2}\ron$.
For a given $d$ the inference seems to remain effectively bounded, and for large $\ron,\roff$ the inference differs from the true values by approximately the diagonal distance on the plane.

This investigation suggests that $d=8$ is roughly on the 25\% relative error threshold for rates approaching 3. 
To tackle the count data from Fig.~\ref{fig:PL}(c) as before, we therefore choose $d=16$ and the results are presented in Fig.~\ref{fig:dtmc-interval_inference}.
The inference is quite accurate, but not as good as the CTMC model.
Increasing $d$ will improve the accuracy of this inference at the cost of increased time to recursively compute the discrete convolutions.

\section{Inferring the state}%
\label{sec:inferring_the_state}

An obvious extension of the work presented above is to determine the underlying state behind a data point given all the data observed.
In the case where the switching is happening many times per detector interval, this task becomes finding what fraction of the interval was spend in a given state.
Here, we illustrate how to solve this task for the case of the single interval model.
In this model the system has a fixed state throughout the detector interval and the task is to determine the state of interval $k$ making use of all observed data, but without knowledge of $\lambda$, $\mu$, $\alpha_1$, or $\beta_1$. 
Specifically, the task is to determine $P(s_k=a|\vec{c}\,I_1)$, where $a$ is either 0 or 1.
For a continuous switching model the posterior distribution will be of the proportion of time spent in a given state underlying each data point.

In order to do the inference we add $\lambda$, $\mu$, $\alpha_1$, $\beta_1$ and the rest of the states and marginalise over them. 
An application of Bayes' rule then yields,
\begin{align}
    P(s_k\!=\!a|\vec{c}\,I_1) =\frac{1}{\mathcal{N}}\int\!\! d\alpha_1 d\beta_1 d\lambda d\mu \sum_{\vec{s}\neq s_k} &
    P(\vec{c}\,|s_1\ldots s_{k-1}\,(s_k\!=\!a)\,s_{k+1}\ldots s_n\,\Omega_1) \nonumber\\
     & \times P(s_1\ldots s_{k-1}\,(s_k\!=\!a)\,s_{k+1}\ldots s_n|\Omega_1)
\end{align}
where $\Omega_1=\alpha_1\,\beta_1\,\lambda\,\mu\,I_1$ as before and the prior probability distributions where taken as constant with the normalisation $\mathcal{N}$ to be determined at the end. 
The summation is over all states $s_j$ where $j\ne k$ denoted in short as $\vec{s}\neq s_k$.
Expanding the last term in temporal order and making use of the independencies in \eqref{eq:I1indep-steps} and \eqref{eq:I1indep-counts} leads to,
\begin{align}
    P(s_k\!=\!a|\vec{c}\,I_1) =\frac{1}{\mathcal{N}}\int\!\! d\alpha_1 d\beta_1 d\lambda d\mu \sum_{\vec{s}\neq s_k} 
        \prod_{t=1}^N P(c_t\,s_t\,|s_{t-1}\,\Omega_1)P(s_0|\Omega_1).
\end{align}

The probability can be efficiently computed using the matrices $R_t$ \eqref{eq:Rmat} and $\vec{D}_0$ \eqref{eq:Dvec}, and either $R_k^{(0)}$ or $R_k^{(1)}$ for $a=0$ or $a=1$ respectively, where
\begin{align}
    R_k^{(0)} = \begin{bmatrix}
        R_{k,00} & R_{k,01} \\
      0 & 0  
  \end{bmatrix} \quad
    R_k^{(1)} = \begin{bmatrix}
      0 & 0 \\
      R_{k,10} & R_{k,11}
  \end{bmatrix} \;
\end{align}
and $R_{k,ij}=P(c_k\,s_{k}\!=\!i|s_{k-1}\!=\!j\,\Omega_1)$ giving the final result
\begin{align}
    P(s_k\!=\!a|\vec{c}\,I_1) =\frac{1}{\mathcal{N}}\int\!\! d\alpha_1 d\beta_1 d\lambda d\mu\; 
    [1\; 1] R_n\cdots R_{k+1}R_k^{(a)}R_{k-1}\cdots R_1 \vec{D}_0.
\end{align}

\begin{figure}
    \includegraphics[width=159.4mm]{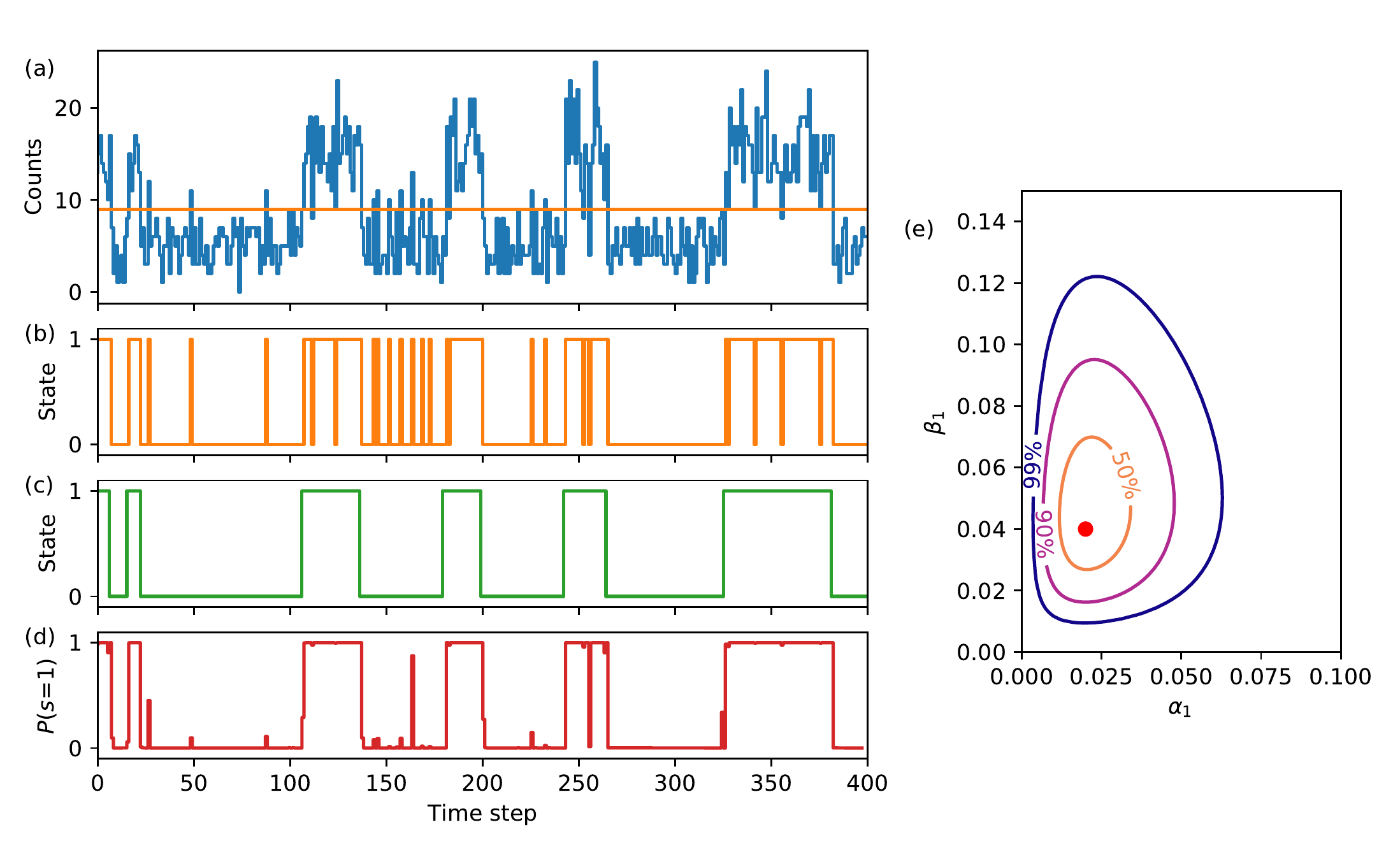}
	\caption{
		(a) Simulated count data from a DTMC single-step model.
        (b) State at each time interval determined by the threshold shown in (a).
        (c) The true state from the simulation.
        (d) Probability the state of an interval was `1' inferred from the entire data sequence.
        (e) The posterior probability of $\alpha_1$ and $\beta_1$ given the same data.
        Note that as more data is observed the credible regions converge of the true values of $\alpha_1$ and $\beta_1$ and the inference of the state becomes much more accurate.
    }
    \label{fig:state_inference}
\end{figure}

To demonstrate the algorithm in action, data was generated and analysed assuming known $\mu$ and $\lambda$, by both a threshold and the full inference and is shown in Fig.~\ref{fig:state_inference}. 
Determining the state by a threshold is shown in Fig.~\ref{fig:state_inference}(b) with the result being either state `0' or state `1'. 
Using the inference based on the whole data yields the probability distribution over the state. 
Fig.~\ref{fig:state_inference}(d) plots the probability that the state is `1', $P(s_k\!=\!1)$, for each data point $k$. 
Note how small fluctuations past the threshold do not significantly affect the probability of the state and are consequently filtered out.
Additionally, the probability of state can convey the confidence in the state assignment---a value of 0.5 means it is equally likely to be either state, this information is lost when a threshold is applied.

\section{Conclusions}

We have derived efficient methods for obtaining the posterior distribution of blinking and emission rates given observation of accumulated detector counts for several blinking models. 
The methods do not require approximations to make the calculation tractable, and make use of all the observed data; the main fitting task is the selection and assumptions that go into the models.
Moreover, the advantage of obtaining the posterior distribution is that rigorous error bounds can be placed on the inferred parameters for example by use of credible regions.
We have also demonstrated how to use the data to determine the underlying state, and unlike a threshold technique the result is automatically carries error bounds in the form of a probability distribution.
It would be an interesting extension to compare this application with other proposals for directly extracting the state e.g. \cite{2014gammelmark043839,watkins2005detection,2010taylor164}.
Though it should be noted that since the rates and the states are dependent on each other the analysis cannot decouple into first determining the states, then from those determining the rates of transitions.

In general we expect the methods presented to apply to a wide variety of hidden Markov models, far beyond the blinking of quantum emitters.
For systems where there is a well-defined state time, the single and multi-step models will be a close fit depending on how the data is accumulated. 
These models are then capable of determining the model parameters for high or low switching probabilities.
In systems that can switch at any time the CTMC model is a better fit, and though it requires the numerical evaluation of an integral as part of the routine we have found that this is not a hindrance.
In these type of systems the stepped models can be thought of as approximations that are more efficient for evaluation when the switching rates are low.
Similarly we expect the methods to readily generalize to more complex Markov models with more states and other methods of indirect observation.

\section*{Acknowledgments}
We would like to thank Yuval Sanders for pointing out the integral solution by Jaynes.
Jemy Geordy acknowledges that the work was done while holding the International Macquarie University Research Excellence Scholarship (``iMQRES MRES") for Master of Research (Allocation number 2018116).
Lachlan Rogers is the recipient of an Australian Research Council Discovery Early Career Award (project number DE170101371) funded by the Australian Government. 
This research was funded in part by the Australian Research Council Centre of Excellence for Engineered Quantum Systems (Project number CE170100009). 

\section*{References}

\bibliographystyle{unsrt}
\bibliography{manuscript}

\end{document}